\documentclass[12pt,preprint]{aastex}

\slugcomment{Accepted for Publication in AJ in December 2006}

\shorttitle{Continued Brightening of Eta Car}
\shortauthors{Martin et al.}

\begin{document}


\title{The Chrysalis Opens?  Photometry from the Eta Carinae HST Treasury Project, 2002--2006\altaffilmark{1,2}}

\revised{30 Aug 2006}

\author{J.\ C.\ Martin\altaffilmark{3}}
\author{Kris\ Davidson}
\and
\author{M.\ D.\ Koppelman}
\affil{University of Minnesota Astronomy Department\\Minneapolis, MN 55455}

\altaffiltext{1}{This research was conducted as part of the Eta
  Carinae Hubble Space Telescope Treasury project via grant
  no. GO-9973 from the Space Telescope Science Institute.  The HST is
  operated by the Association of Universities for Research in
  Astronomy, Inc., under NASA contract NAS5-26555.} 
\altaffiltext{2}{Some of the data presented in this paper were
  obtained from the Multi-mission Archive at the Space Telescope
  Science Institute (MAST). STScI is operated by the Association of
  Universities for Research in Astronomy, Inc., under NASA contract
  NAS5-26555. Support for MAST for non-HST data is provided by the
  NASA Office of Space Science via grant NAG5-7584 and by other grants
  and contracts.} 
\altaffiltext{3}{Now at the University of Illinois Springfield, Springfield, IL 62703}



\begin{abstract}
During the past decade $\eta$ Car has brightened markedly, 
possibly indicating a change of state.  Here we summarize photometry 
gathered by the Hubble Space Telescope as part of the HST Treasury 
Project on this object.  Our data include STIS/CCD acquisition images, 
ACS/HRC images in four filters, and synthetic photometry in 
flux-calibrated STIS spectra.  The HST's spatial resolution allows 
us to examine the central star separate from the bright circumstellar 
ejecta.  {\it Its apparent brightness continued to increase briskly 
during 2002--06, especially after the mid-2003 spectroscopic event.\/}  
If this trend continues, the central star will soon become brighter 
than its ejecta, quite different from the state that existed only 
a few years ago.  One precedent may be the rapid change observed
in 1938--1953.  We conjecture that the star's mass-loss rate has 
been decreasing throughout the past century.      
\end{abstract}




\section{Introduction}      

Eta Carinae's photometric record is unparalleled among 
well-studied objects, especially since it has been near or exceeded
the classical Eddington limit during the past two or three 
centuries.  From 1700 to 1800 it gradually brightened from 4th to 
2nd magnitude, and then experienced its famous Great Eruption or 
``supernova impostor event'' beginning about 1837.  For twenty years 
it was one of the brightest stars in the sky, rapidly fluctuating 
between magnitudes 1.5 and 0.0, briefly attaining $V \approx -1.0$.
After 1858 it faded below 7th magnitude, presumably enshrouded in 
the nascent Homunculus nebula.  Subsequent behavior, however, 
has been more complex than one might have expected.    
A mysterious secondary eruption occurred in 1887--1900; then the 
apparent brightness leveled off around $m_{pg} \approx 8$ for about 
40 years, followed by a rapid increase in 1938--53;  after that 
it brightened at a fairly constant rate for another 40-year 
interval, and most recently the rate accelerated in the 1990's.   
Some, but not all, of the secular brightening can be attributed to 
decreasing obscuration as the Homunculus nebula expands.  However, in
truth this is more complex than it appears.  
The star's physical structure has been changing in a decidedly
non-trivial way which is, at best, only dimly understood.  For 
historical and observational details see  \citet{dh97,frew05,dvauc52,
oconnell56,feinstein67,feinstein74,aavso,kdetal99a,kdetal99b,
vangenderen99, sterken99,phot1}.

Spectroscopic changes have occurred along with the brightness
variations.  The 5.5-year spectroscopic/photometric cycle
\citep{gaviola53,zanella84,whitelock94,whitelock04,damineli96,phot1}
is not apparent in data obtained before the 1940's
\citep{feast01,rmh05}.  Brief ``spectroscopic events'' 
marking the cycle are most likely mass-ejection or 
wind-disturbance episodes, probably regulated by a 
companion star \citep{zanella84,kd99,smith03,heii}.  
At visual wavelengths, the associated ephemeral brightness 
changes represent mainly emission lines in the stellar wind, 
while the longer-term secular brightening trend involves 
the continuum \citep{phot1,martinconf}.  \citet{rmh05},
\citet{halpha05}, and \citet{kd05} have speculated 
that the four obvious disruptions in the photometric record --
c.\ 1843, 1893, 1948, and 2000 -- might indicate a quasi-periodicity 
of the order of 50 years.\footnote{
So far as we know, this idea was first voiced by Humphreys at two 
meetings in 2002, but it did not appear in the published proceedings}
In any case the star has not 
yet recovered from its Great Eruption seen 160 years ago.  

The Hubble Space Telescope (HST) Treasury Program for $\eta$ Car was 
planned specifically to study the 2003.5 spectroscopic event.  We
employed the Space Telescope Imaging Spectrograph (STIS) 
and Advanced Camera for Surveys (ACS), following earlier STIS 
observations that began in 1998 \citep{etatp04}.   Fortuitously, the 
STIS data almost coincide with a rapid secular brightening which 
began shortly before 1998 (see Section 4 below).  Those and the ACS 
images are of unique photometric value for at least two reasons:
\begin{enumerate}
\item{At visual wavelengths, normal ground-based observations have 
  been dominated by the surrounding Homunculus ejecta-nebula, which,
  until recently, appeared much brighter than the central star and 
  which has structure at all size scales from 0.1 to 8 arcseconds.  
  {\it So far, only the HST has provided well-defined measurements 
  of just the central star.\/}\footnote{
    At least this is true for visual and UV wavelengths.  The
  near-infrared photometry reported by \citet{whitelock94} and
  \citet{whitelock04} may be strongly dominated by the central star.
  Those observations probably represent free-free emission in the wind
  at larger radii than the visual wavelength data.  They show both the
  spectroscopic events and the brightening trend better than other
  ground-based measurements.} 
  The Homunculus is primarily a reflection 
  nebula, but the Homunculus/star brightness ratio has changed 
  substantially. During 1998-99, for instance, the star nearly tripled 
  in apparent brightness while ground-based observations showed only 
  about a 0.3-magnitude brightening of Homunculus plus star 
  \citep{kdetal99b}.  This rather mysterious development is known 
  from HST/STIS and HST/ACS data. }

\item{Numerous strong emission lines perturb the results for standard 
  photometric systems.  H$\alpha$ and H$\beta$ emission, for example, 
  have equivalent widths of about 800 and 180 \mbox{\AA} respectively 
  in spectra of $\eta$ Car.  Broad-band $U$, $B$, $R$, and $I$ magnitudes, 
  and most medium-band systems as well, are therefore poorly 
  defined for this object.  Photometry around 5500 \mbox{\AA}, e.g. 
  broad-band $V$, is relatively free of strong emission lines, but 
  transformations from instrumental magnitudes to a standard system 
  require the other filters \citep{kdetal99b,sterken01,vangenderen03}.
  This difficulty is somewhat lessened for HST observations restricted
  to the central star, whose spectrum has fewer emission lines
  than the bright ejecta;  and some of the HST/ACS filters are
  fairly well-adapted to the case.  At any rate the STIS and ACS
  data appear to be stable and internally consistent.  Detector and 
  filter systems used in most ground-based work, on the other hand,
  require fluctuating instrumental and atmospheric corrections, and
  do not give any major advantage for this object. }
\end{enumerate}
In this paper we present the complete set of photometric data gathered 
for the $\eta$ Car HST Treasury Project.  The central star has brightened,
especially in the UV, since the HST results described by \citet{phot1}.  
We report three types of later measurements:
\begin{enumerate}
  \item{The star's brightness in acquisition images made with STIS 
  before that instrument's failure in early 2004.  These images represent 
  a broad, non-standard wavelength range from 6500 to 9000 {\AA}.}   
  \item{The star's brightness in ACS/HRC images made in four filters
  (F220W, F250W, F330W, \& F550M) from October 2002 to the present.}
  \item{Synthetic photometry in flux-calibrated STIS spectra.}
\end{enumerate}

After presenting  the data below, we briefly discuss the observed
trends and what they bode for the near future of $\eta$ Car.  Our
principal reason for reporting these data now is that the Treasury
Program observations have been completed;  future HST/ACS observations
are possible but not assured.  These last Treasury Program
observations are essential for demonstrating the secular trend in
brightness (see Section \ref{trend}), that these brightness changes
suggest a fundamental change in the state of the star (see Section
\ref{discussion}) and what the near future may hold for $\eta$ Carinae
(\ref{predictions}).

\section{Data}     

\subsection{STIS Acquisition Images}

\begin{figure}
\includegraphics[angle=90,scale=0.7]{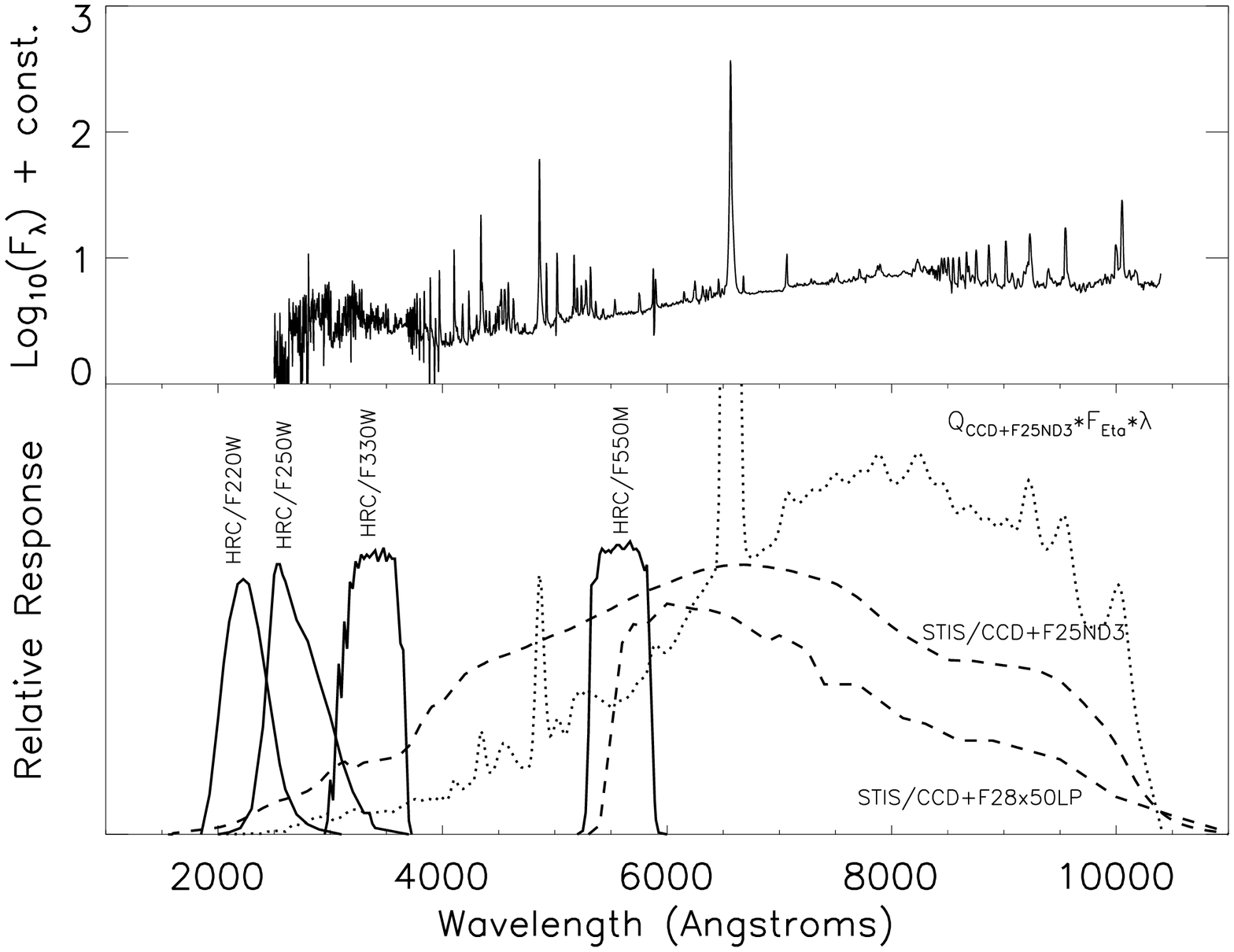}
\caption{Photometric response functions.  The top panel shows the
  relative spectral flux from the central star of $\eta$ Carinae.  The
  bottom panel shows the total relative response of each CCD and
  filter combination used in this study on the same wavelength scale
  as the top panel.  For plotting purposes the curves are not
  representative of relative responses between filters.  STIS filters
  are plotted with a dashed line and ACS/HRC filters are plotted with
  a solid line.  The dotted line represents the product of the STIS
  CCD+F25ND3 response curve and the photon flux from the central
  star.} 
\label{filterfig}
\end{figure}

Each set of STIS observations included a pair of acquisition images, 
which are 100 x 100 pixel sub-frames ($5{\arcsec} \times 5{\arcsec}$) 
centered on the middle row and column of the CCD
\citep{clampin96,downes97,kimquijano03}.  The STIS acquisition images
were taken with a neutral density filter (F25ND3) which, combined with
the CCD response, covered the wavelength range 2000--11000 {\AA}.
Since the star's apparent color is moderately red, these images were
dominated by fluxes at wavelengths 6500--9000 {\AA}.  Fig.
\ref{filterfig} shows that although most of the measured flux comes
from the continuum, several prominent emission features including
H$\alpha$ contributed to the measured brightness.   We measure the star's
brightness with in radius $R$ in the following manner:  If $f(x,y)$
denotes the flux level in an image where the star was centered at $x_0$,
$y_0$, we integrate the product of $w(x-x_0,y-y_0) f(x,y)$, where $w$ 
is a radially symmetric weighting function of the form $(1-r^2/R^2)$,$r<R$.
In effect $w$ is a ``parabolic virtual field aperture.'' For the STIS
acquisition images we chose $R = 0.3{\arcsec}$ (6 pixels).  A detailed
discussion of these reduction procedures, including a non-standard bias
level correction which we applied, is given in \citet{phot1}.

Initially we expected to add a few more points of acquisition
photometry after 2004.5, but the failure of the STIS in August 2004 
curtailed our observing plans.  Thus we have only two additional 
STIS data points to report (Table \ref{acqdata}) beyond those 
given in \citet{phot1}.  They are useful, however, concerning  
the end of the post-event recovery (see Sections \ref{trend} and
\ref{discussion} below).

\subsection{HST ACS}
HST ACS/HRC observations of $\eta$ Car were obtained for the Treasury
Project beginning in October 2002.  We have also examined publicly
available data from HST proposal 9721\footnote{``The Kinematics and
  Dynamics of the Material Surrounding Eta Carinae,'' B. Dorland,
  principal investigator} and HST proposal 10844\footnote{``Following
  Eta Carinae's Change of State,'' K. Davidson, principal investigator}.  
The bias-corrected, dark-subtracted, and flat-fielded data were obtained 
from the Space Telescope Science Institute via the Multi-Mission Archive 
(MAST) \citep{acscal}.\footnote{\tt http://archive.stsci.edu} 

Treasury Program ACS/HRC images were taken in four filters that cover 
near-UV to visual wavelengths (Fig. \ref{filterfig}):
\begin{itemize}
\item{HRC/F220W and HRC/250W:  These near-UV filters sample the 
  ``\ion{Fe}{2} forest'' \citep{cassatella79,altamore86,viotti89}, 
  whose opacity increases dramatically during a spectroscopic 
  event \citep{kdetal99c,ironcurtain}.}
\item{HRC/F330W:  This filter includes the Balmer continuum
  in emission, supplemented by various emission lines.   
  It also attained the best spatial resolution
  among the observations reported here.}
\item{HRC/F550M:  With a medium-width (not broad) bandpass, this
  filter samples the visual-wavelength continuum flux with only 
  minor contamination by emission features.}
\end{itemize}

The brightness of the central star was measured using the same 
0.3$\arcsec$ ($\sim$ 10 ACS/HRC pixels) weighted virtual aperture 
used for the STIS acquisition images.  CCD flux values were converted 
to the STMAG system \citep{stmag} using the keywords provided in the 
MAST archive's FITS headers.  An aperture correction \`{a} la   
\citet{acscal}, calculated from observations of the white 
dwarf GD~71 in each of the filters (Table \ref{acscaltab}), was  
applied to the measurements (Table \ref{acsdata}).  ACS fluxes 
and magnitudes measured prior to MJD 52958 (2003.87) can be found 
in our first paper \citep{phot1}.  {\it Caveat:\/} We did not apply 
the aperture corrections to the magnitudes in that paper, but we 
have done so in the plots shown here.

The F220W and F250W filters have known red leaks \citep{acshandbook}
that can affect photometry of red sources.  We
convolved extracted STIS spectra (see Sec. \ref{stisphot}) with the
response function for those filters and the ACS/HRC.  In the case of
the central star of $\eta$ Car, the flux redward of 4000{\AA} in the
F220W and F250W filters contributed only about 0.25\% and 0.06\%
respectively, insignificant compared to other sources of error. 

\subsection{STIS Synthetic Photometry\label{stisphot}}

Originally the ACS/HRC images were meant to supplement the STIS
spectra.  After the untimely demise of the STIS, however, the ACS/HRC
became the most suitable mode for observing $\eta$ Car with HST.  
This presented a problem of continuous monitoring in the same band 
passes over the entire program, since there were no ACS/HRC images 
prior to 2002.78 while the STIS data ended at 2004.18. 

ACS photometry can be synthesized from the flux calibrated STIS
data, since nearly every grating tilt was observed during each STIS
visit (Table \ref{stisspectra}).  The spectra were extracted with a
weighted parabolic cross dispersion profile similar to the virtual
aperture used to measure the ACS/HRC images, convolved with the 
published filter functions (Fig \ref{filterfig}), and integrated.  Because 
STMAG is computed from flux density, the integrated fluxes were 
divided by the effective band passes of each filter 
(see Table \ref{acscaltab}). 

The effective aperture for the extracted STIS spectral data is not a
rotated parabola but a parabolic cylinder having the width of the slit
(0.1\arcsec).  To correct for the difference in aperture as well as
the difference in instrumental PSF, slit throughput, and the extraction 
height, we converted the STIS spectral fluxes to the ACS/HRC flux scale 
using suitable correction factors (Table \ref{acscaltab}).  Those 
factors were computed by comparing the results from ACS/HRC images 
and photometry synthesized from STIS data obtained at time MJD 52683.  
The resulting synthetic ACS/HRC photometry is given in Table \ref{fakeacs}.

Plots of these various data will be discussed in Sections \ref{trend}
and \ref{discussion} below.

\section{Distribution of Surface Brightness in the Homunculus}  

Since HST and ground-based photometry have shown different rates 
of change for different-sized areas, it is useful to view the 
spatial distribution of the brightness.  For this purpose we have 
used three ACS images made with filter F550M at $t =$ 2004.93, 
2005.53, and 2005.85.   Together these give a reasonably valid  
picture of the average visual-wavelength appearance during 2005. 

\begin{figure}
\plotone{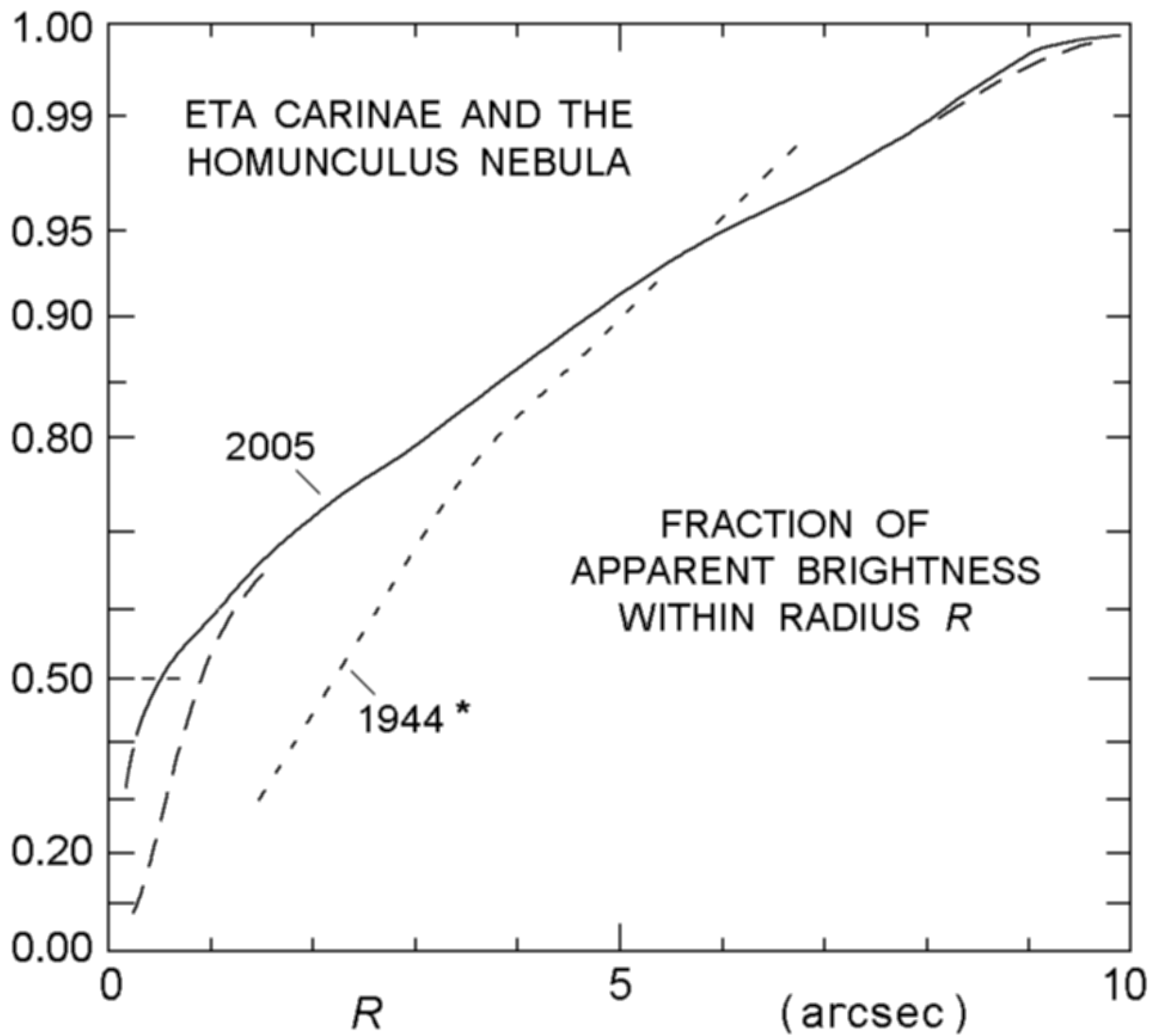}
\caption{Fraction of total visual-wavelength brightness
    originating within projected radius $R$, based on
    HST/ACS images described in the text.  The solid
    curve represents the appearance with high spatial
    resolution, while the nearby dashed curve shows
    the result of Gaussian blurring with FWHM 0.8{\arcsec},
    roughly equivalent to atmospheric ``seeing.''  Another, 
    lightly dashed curve refers to Gaviola's photographs 
    made in 1944 \citep{gaviola50,kdmtr75}, with $R$ 
    multiplied by 1.6 to compensate for nebular expansion 
    between 1944 and 2005. [ 1.6 = $(2005 - 1843) / 
    (1944 - 1843)$, where 1843 is the characteristic date
    of ejection.]} 
\label{brt-distr-1}
\end{figure}

Fig. \ref{brt-distr-1} shows the fraction of apparent brightness
within projected radius $R$ measured from the central star.  The 
solid curve represents the HST data, while a companion dashed curve 
incorporates Gaussian blurring with FWHM = 0.8{\arcsec}, 
simulating ground-based photometry with fairly good atmospheric 
conditions.  Half the total light originates within 
$R \lesssim 0.5{\arcsec}$ -- which is very different from 
$\eta$ Car's appearance a few decades ago \citep{gaviola50,adt53,
gn72,kdmtr75,vgt84}.  Fig.\ \ref{brt-distr-1} also shows a curve 
based on photographs that Gaviola obtained in 1944 \citep{gaviola50,
kdmtr75}, with a magnified spatial scale to compensate 
for subsequent expansion;  even allowing for 
mediocre ``seeing'' and other uncertainties, the degree of central
condensation was obviously less then. Before 1980 the central star 
accounted for less than 10\% of the total apparent brightness;  
now its fraction has grown to about 40\% and continues to increase.  

\begin{figure}
\plotone{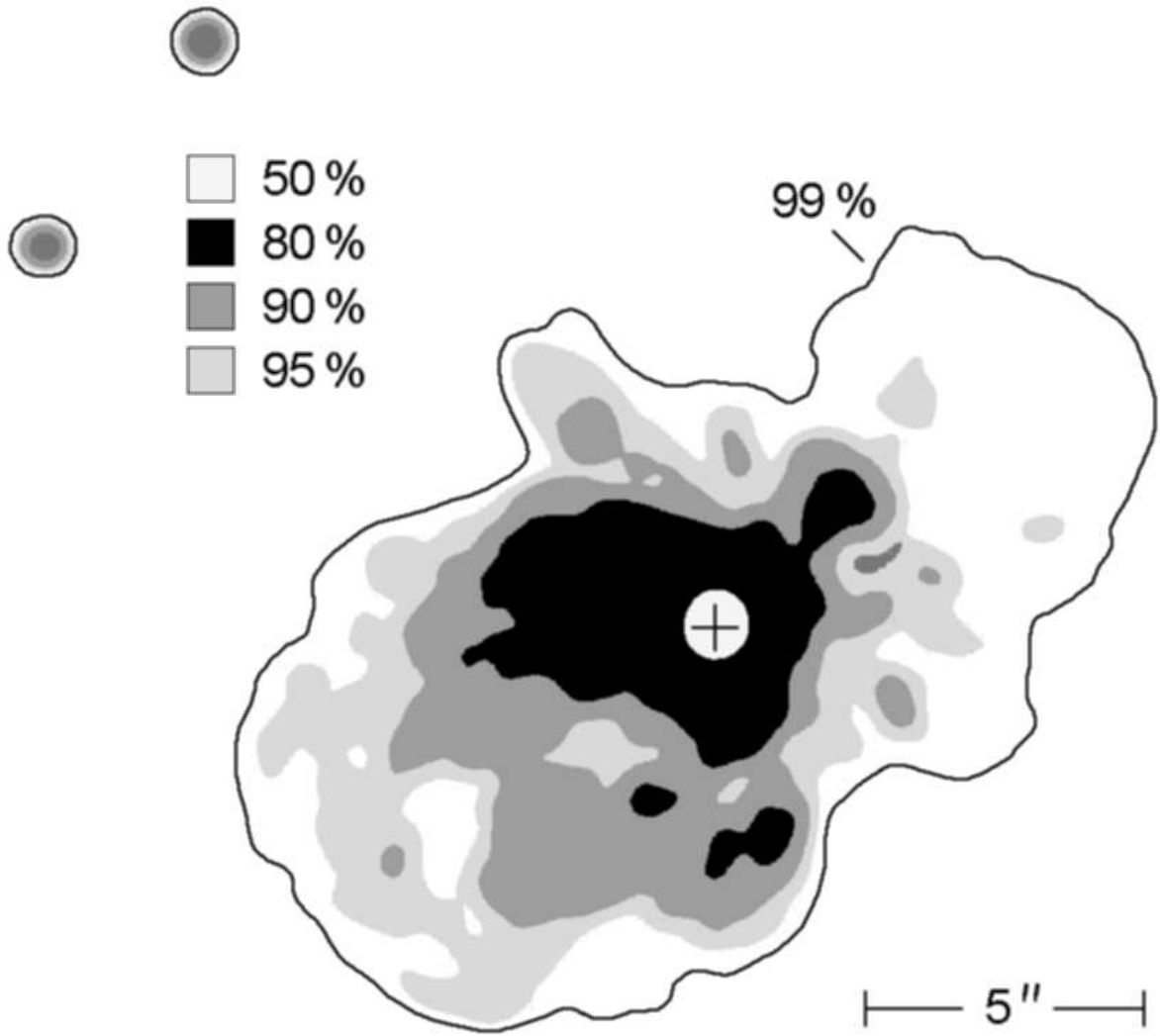}
\caption{Visual-wavelength isophotes in the Homunculus
    nebula.  In order to simplify the contours, the image 
    has been blurred (convolved) with a circular Gaussian 
    with FWHM = 0.5{\arcsec}.  The isophote levels were chosen 
    to enclose specified fractions of the total integrated 
    brightness;  for example, 50\% of the apparent light comes 
    from within the innermost, roughly circular isophote.  
    Relative to the central maximum in the blurred image, 
    the isophotes marked 50\%, 80\%, 90\%, 95\%, and 99\% 
    have intensities  0.60, 0.080, 0.0034, 0.00149, 0.00076, 
    and 0.00028, respectively.  North is at the top of
    this figure, east is to the left, and the two circular
    objects in the upper left corner are stars in the
    images. {\it Caveat:\/} This map represents conditions 
    in the year 2005;  the degree of central condensation 
    is progressively increasing. }
\label{brt-distr-2}
\end{figure}

Fig. \ref{brt-distr-2} is a map of the surface brightness, based 
on the HST/ACS F550M data mentioned above.  In order to produce 
simple well-defined isophotes, we have Gaussian-blurred 
the image using FWHM 0.5{\arcsec}.  Apart from the 
central star and nearby compact ejecta, most of the light 
comes from a comma- or crescent-shaped region about
5{\arcsec} across, marked by the 80\% isophote in the
figure.  Presumably the high intensity in this area results 
from strong forward scattering by dust grains in that part 
of the southeast Homunculus lobe.  Surface brightnesses 
in the outer lobe regions, on the other hand, are fainter 
than the 50\% contour by factors typically between 100 and 
200.   About half of the projected area of the Homunculus 
provides only 5\% of the total visual-wavelength brightness.

\section{The Eight-year Trend\label{trend}}  

Fig.\ \ref{plot1} shows the main HST photometric data on the central
star;  the most significant result is a secular brightening trend
superimposed  on the  5.5-year pattern of spectroscopic events.  The
latest observations are essential in this regard,  because the data
reported earlier by \citet{phot1} ended before the star had emerged
from the 2003.5 event and we could not be sure of the long term trend.
From 1999 to 2006 the average trend was about 0.15 magnitude per year
at visual  wavelengths and 0.19 mag yr$^{-1}$ at 2200 {\AA} -- much
faster than the rate of $\sim$ 0.025 mag yr$^{-1}$ recorded for the
Homunculus plus star from 1955 to 1995 \citep{kdetal99b,martinconf}. 

The Weigelt condensations northwest of the central star 
\citet{speckle86} have {\it not\/} brightened rapidly.  Located 
in the equatorial plane only about 800 AU from the star\footnote{
For conversions between apparent and linear size scales we assume that 
$\eta$ Car's distance is 2300 pc \citep{dh97}.
}, their light
is intrinsic emission with some reflection \citep{kdfos95}. 
Fig.\ \ref{knotd} shows the brightness of ``Weigelt D,'' measured in 
the same way as the star but centering the virtual aperture at offset 
location $r = 0.25{\arcsec}$, position angle 336$^{\circ}$.  
The absence of a strong secular trend is
significant in the following way.  Extrapolating the recent trend of
the star/ejecta brightness ratio back to the mid-1980's, one would  
expect that the star should have been fainter than each 
of the Weigelt blobs at that time.  But this is contradicted
by early speckle observations \citep{speckle86,speckle88};
{\it therefore the star cannot have brightened at the present-day rate
through the entire 20-year interval.\/} Moreover, the earliest HST/FOS
spectroscopy in 1991 \citep{fos95} and the ultraviolet spectra of the
star plus inner ejecta obtained with the International Ultraviolet
Explorer (IUE) from 1979 to 1990 show absolute fluxes which, though
uncertain, appear comparable to both the speckle observations and  the
1998 STIS results.  These facts imply that the central star's brightening
rate was relatively modest from 1980 until sometime in the 1990's.  We
suspect that the present-day rate began in 1994--97, when
ground-based  photometry showed unusual behavior (see, e.g., Fig.\ 2
in \citet{kdetal99b}). 

The last F330W and F550M observations in Fig. \ref{plot1} confirm 
the sudden 0.2 magnitude increase observed at La Plata in late 2005
\citep{laplata}.

\section{Discussion\label{discussion}}  

The observed brightening of $\eta$ Car is not easy to explain.  
It cannot signify a major increase in the star's luminosity,
because that would exceed the Eddington limit, producing a giant   
eruption.  It cannot be a standard LBV-like eruption;  in that case 
the energy distribution should have shifted to longer wavelengths, the
Balmer emission lines should have decreased, and the spectrum should 
have begun to resemble an A- or F-type supergiant \citep{lbv94}.  
In fact, qualitatively the star's spectrum has changed little in the 
past decade, and it has become bluer, not redder.\footnote{ The
  change in color is modest, however, too small to confidently
  quote here.  Dust near $\eta$ Car has long been known to have an
  abnormally small reddening/extinction ratio, see \citet{dh97},
  \citet{fos95}, and refs.\ cited therein.}  

The most obvious remaining explanation involves a change in the
circumstellar extinction,  which, in turn, probably requires a   
subtle change in the stellar wind.  Mere ``clearing of the dust''  
-- i.e., motion of a localized concentration of dusty ejecta -- cannot
occur fast enough \citep{kdetal99b}.  Therefore one must consider
either destruction of dust grains, or a decrease in the formation of
new dust, or both; and, if these account for the observations, why
should they happen now? 

\subsection{Dust Near the Star} 

The hypothetical decreasing extinction probably occurs within 2000 AU
($\sim$ 1{\arcsec}) of the star, and preferably closer, because:
\begin{enumerate}
\item{In various observations between 1980 and 1995, the star 
 did not appear as bright as expected relative to the Weigelt
 blobs;  the discrepancy was a factor of the order of 10, based 
 on simple theoretical arguments \citep{dh86,fos95}.  Evidently,
 then, our line of sight to the star had substantially larger   
 extinction at visual wavelengths, even though its projected 
 separation from the blobs was less than 0.3{\arcsec}.  The required 
 extra extinction was of the order of 3 magnitudes.  Since then 
 the star has brightened far more than the Weigelt objects have;
 therefore, if this involves localized extinction, its size 
 scale must be a fraction of an arcsec, only a few hundred AU. } 
\item{No known process seems likely to destroy dust more than 
 2000 AU from the star in a timescale of only a few years.}
\item{Ground-based photometry and HST images have shown only a 
 modest, fraction-of-a-magnitude increase in the brightness of 
 the large-scale Homunculus lobes during the past decade 
 \citep{laplata,phot1}. } 
\end{enumerate}

Dust grains should condense in $\eta$ Car's wind at a distance of 
200--600 AU, 2 to 10 years after the material has been ejected.\footnote{
Here, lacking a specific dust-formation model for the unusual case of 
$\eta$ Car, we suppose that appreciable grain condensation begins in the 
outward flowing material at the location where the equilibrium grain 
temperature is around 1000 K.  This is a fairly conventional assumption
and the precise choice of temperature has little effect on our reasoning.
The quoted time-after-ejection assumes typical ejecta speeds of 
200--700 km s$^{-1}$.}  Since
newly-formed dust moves outward in a timescale of several years, the
circumstellar extinction seen at any time depends partly on the
current dust formation rate.  This, in turn, depends on local wind
density, radiation density, etc., and newly formed hot grains ($T_d >
800$ K) are susceptible to destruction.  The dust column density can
thus be sensitive to small changes in the stellar parameters.
Moreover, the wind is latitude-dependent and our line of sight is
close to the critical latitude where wind parameters can vary rapidly
\citep{smith03}.  All these factors appear suitable for the proposed
explanation. 

On the other hand, near-infrared observations imply that extinction 
within $r < 2000$ AU  has been quite small along most paths outward 
from the star.  In Fig.\ 3 of \citet{cox95}, for instance, the
2--6 $\mu$m flux indicates the high end of the dust temperature 
distribution.  Modeling this in a conventional way, we find   
that less than 5\% of the total luminosity was absorbed and  
re-emitted by inner dust with $T_d > 500$ K during the years 
1973 to 1990 when those observations were made.\footnote{
The measured flux was approximately a power law 
$f_{\nu} \sim {\nu}^{-3.7}$ at wavelengths around 4 ${\mu}$m.
Assuming a typical emission efficiency dependence $Q_{\nu} \sim {\nu}$,
the observed spectral slope can be explained by a grain temperature
distribution $dN/dT \sim T^{-8.7}$.  The result noted in the text is
obtained by normalizing this to match the observed flux around 
4 or 5 $\mu$m and then integrating the total emitted flux at all
wavelengths due to grains above 500 K.}  Therefore, 
{\it our line of sight must be abnormal in order to have a large 
amount of extinction near the star.\/}  In principle one might 
view this as an argument against our proposed scenario, but no 
plausible alternative has been suggested to explain the apparent 
faintness of the central star before 1999 and its ratio to the 
Weigelt blobs \citep{speckle86,dh86,speckle88,fos95,kdetal99b},       

The spatial distribution of dust is probably quite inhomogeneous
near the star.  The Homunculus lobes have a conspicuously ``granular''
appearance;  the equatorial ejecta are clumpy, including the 
Weigelt knots; and stars near and above the Eddington limit 
tend to produce clumpy outflows \citep{shaviv05}.  Consequently 
the radiative transfer problem includes macroscopic 
effects which have not yet been modeled.  If the grain albedo
is sufficient, light may escape mainly by scattering through 
interstices between condensations.  In that case, high-extinction 
lines of sight may be fairly common in the inner region even 
though most of the light escapes along other paths, not 
necessarily radial.       

Incidentally, the near-infrared photometric trends reported by
\citet{whitelock94,whitelock04} are not straightforward
to interpret.  The fairly-constant 3.5 $\mu$m flux, for instance,
represents a complicated mixture of dust formation parameters
and does not necessarily indicate a constant amount of dust;
see comments by \citet{kdetal99b}.

\subsection{The Role of the Stellar Parameters \label{stellarparams}}

If the observed brightening represents a decrease in
circumstellar extinction, the likeliest reason for this
to occur is through some change in the star -- no one has yet proposed
a suitable alternative.  The most relevant stellar parameters are
the radius,  current luminosity, and surface rotation rate, which 
together determine the wind's velocity, density, and 
latitude structure.  All of these may still be changing 
today, 160 years after the Great Eruption;  thermal and 
rotational equilibrium in particular are likely to be 
poor assumptions for the star's internal structure 
\citep{smith03,kd05}. 

As a working hypothesis to explain $\eta$ Car's photometric 
and spectroscopic record in the past 100 years, let us
tentatively suppose that {\it the mass-loss rate is gradually 
decreasing,\/} while the surface rotation rate may be 
increasing.  Historical considerations include:  
\begin{enumerate}
\item{High-excitation \ion{He}{1} emission, now observed
at most times, was consistently absent before 1920 \citep{feast01} 
and probably before 1940 \citep{rmh05}.  If a hot companion
star is present as most authors suppose, then the most obvious 
way to hide or suppress its helium ionization is to immerse 
the entire system in an extremely dense wind -- i.e.,
the primary star's mass-loss rate must have been larger then.  
This idea is far from straightforward \citep{kd99}, but 
so far as we know it is the only qualitative 
explanation yet proposed.  Informally, based on Zanstra-style arguments 
(i.e., assessing the volume emission measure $n_{He} n_e V$ needed to 
absorb all the photons above 25 eV), 
we estimate that a rate of the order of 10 times the
present value, i.e. $\sim 10^{-2}$ $M_\odot$ y$^{-1}$, would 
have been required early in the twentieth century in order to suppress
the helium recombination emission. }      
\item{Twenty years ago the amount of fresh dust, indicated
by the near-infrared flux, appeared consistent with a 
mass-loss rate somewhat above $10^{-3}$ $M_{\odot}$ yr$^{-1}$
\citep{iue86}.  This absorbed only a small fraction of the
luminosity (Section 5.1 above), but the substantially higher
mass-loss rate suspected for earlier times would have 
produced enough hot inner dust to absorb a non-negligible 
fraction. }  
\item{The brightness observed between 1900 and 1940 is rather
mysterious.  Judging from its mass and present-day optical 
thickness, around 1920 the Homunculus (then only half as 
large as it is today) should have had at least 5 magnitudes of 
visual-wavelength extinction;  in a simple model the object 
should have been fainter than 10th magnitude instead of 
$m_{pg} \approx 8$ as was observed. No doubt the inhomogeneities 
mentioned earlier played a role, but no model has been calculated.  
Moreover, why did the brightness remain fairly constant even
though the Homunculus expanded by about 70\% in 1900--1940?  
This interesting problem has received practically no 
theoretical attention.}
\item{\ion{He}{1} emission first appeared, and $\eta$ Car's
brightness suddenly increased, between 1938 and 1953 as we
mentioned in Section 1.  This might conceivably be explained
by a decrease in the wind density;  but \citet{kd05} and
\citet{halpha05} have conjectured that 1940--1950 may 
have been the time when rotation became fast enough to 
produce latitude structure in the wind.  If so, a 
higher-excitation, lower-density zone then developed at 
low latitudes \citep{smith03}. }  
\end{enumerate}

The above points inspire two hypotheses that may explain
the rapid brightening trend shown in Fig.\ \ref{plot1}. 
First, if the mass-loss rate has been decreasing, 
this tends to reduce the column density of 
recently-formed dust along our line of sight.  Meanwhile 
(or alternatively), perhaps the wind's latitude structure 
is continuing to evolve so that its dense zone is now moving 
out of the line of sight.  HST data suggest that our line 
of sight has been fairly close to the critical boundary 
latitude separating the two phases \citep{smith03}.  A 
small increase in surface rotation rate, or some other 
parameter change, might conceivably move the dense zone 
to higher latitudes, decreasing the amount of dust that
forms along our line of sight.  This idea is appealing 
because it suggests a way in which the effective
extinction may be very sensitive to the stellar parameters.   

This problem obviously requires detailed models far beyond
the scope of this paper, combining the star's changing 
structure, its wind, dust formation, and possibly dust 
destruction.    

\subsection{Concerning the 5.5-year Cycle}

\begin{figure}
\plotone{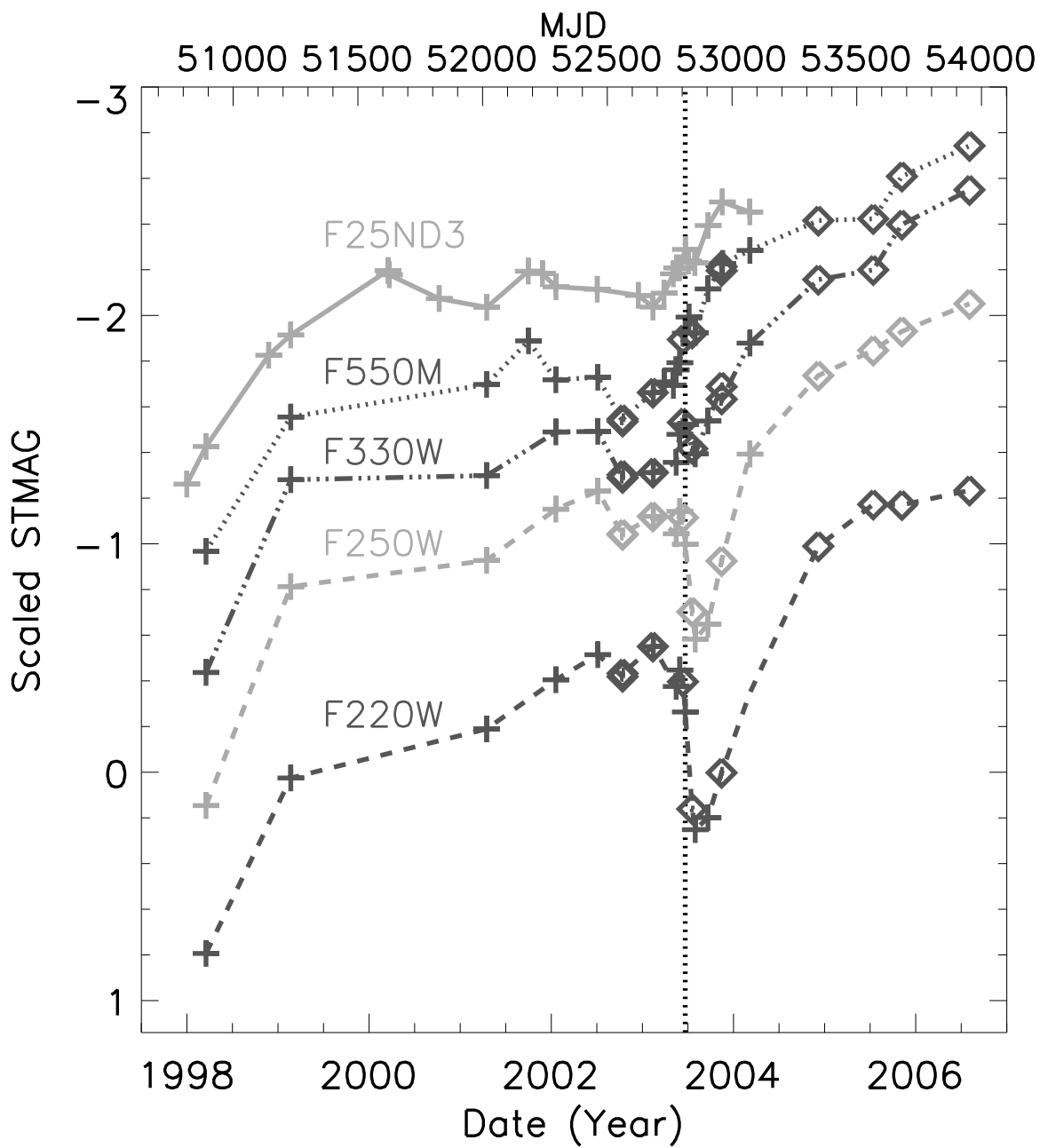}
\caption{Photometry of the central star.  The magnitudes in each filter 
are scaled by arbitrary amounts for plotting them together.  Crosses 
denote data from STIS ACQ images or synthetic photometry derived from STIS 
CCD spectra.  Diamonds denote photometry measured from ACS/HRC data.  The
formal statistical errors are smaller than the size of the symbols used.  The
vertical dotted line marks the time of the spectroscopic event in 2003.}
\label{plot1}
\end{figure}

\begin{figure}
\plotone{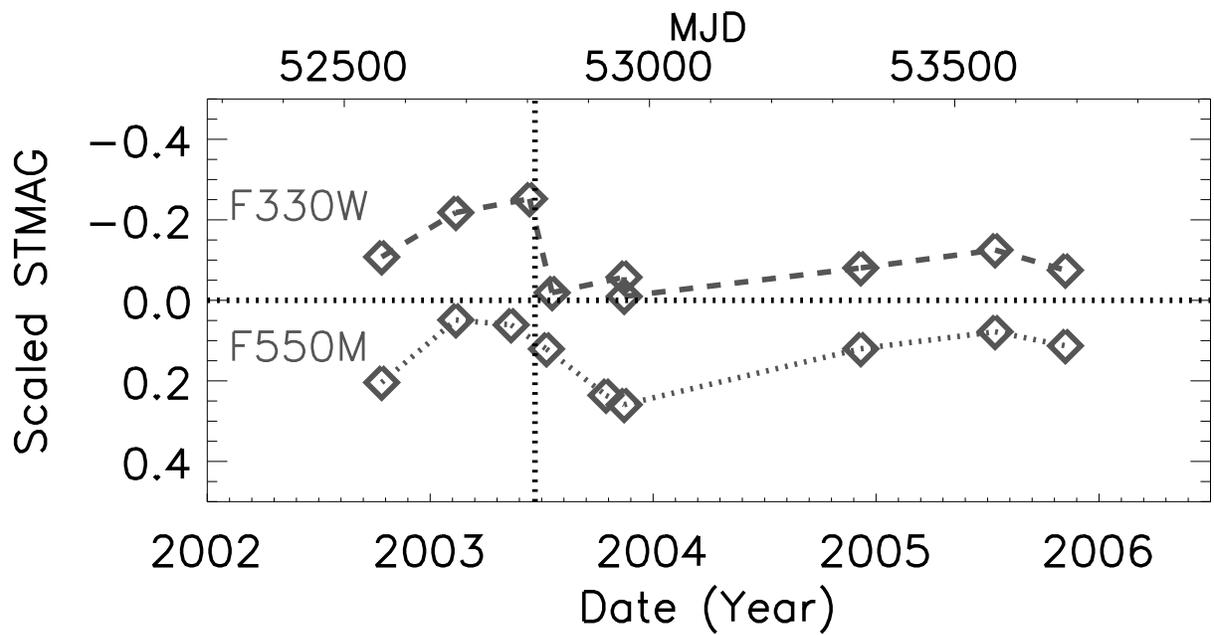}
\caption{ACS measurements of brightness of the equatorial ejecta near
  ``Weiglet D'' ($r = 0.25{\arcsec}$, PA = 335.5$^\circ$). The time of
  the spectroscopic  event in 2003 is marked with a vertical dashed line.
  Various STIS data (unpublished and not plotted here) confirm that no 
  strong secular trend occurred in 1998 -- 2004.} 
\label{knotd}
\end{figure}

\begin{figure}
\plotone{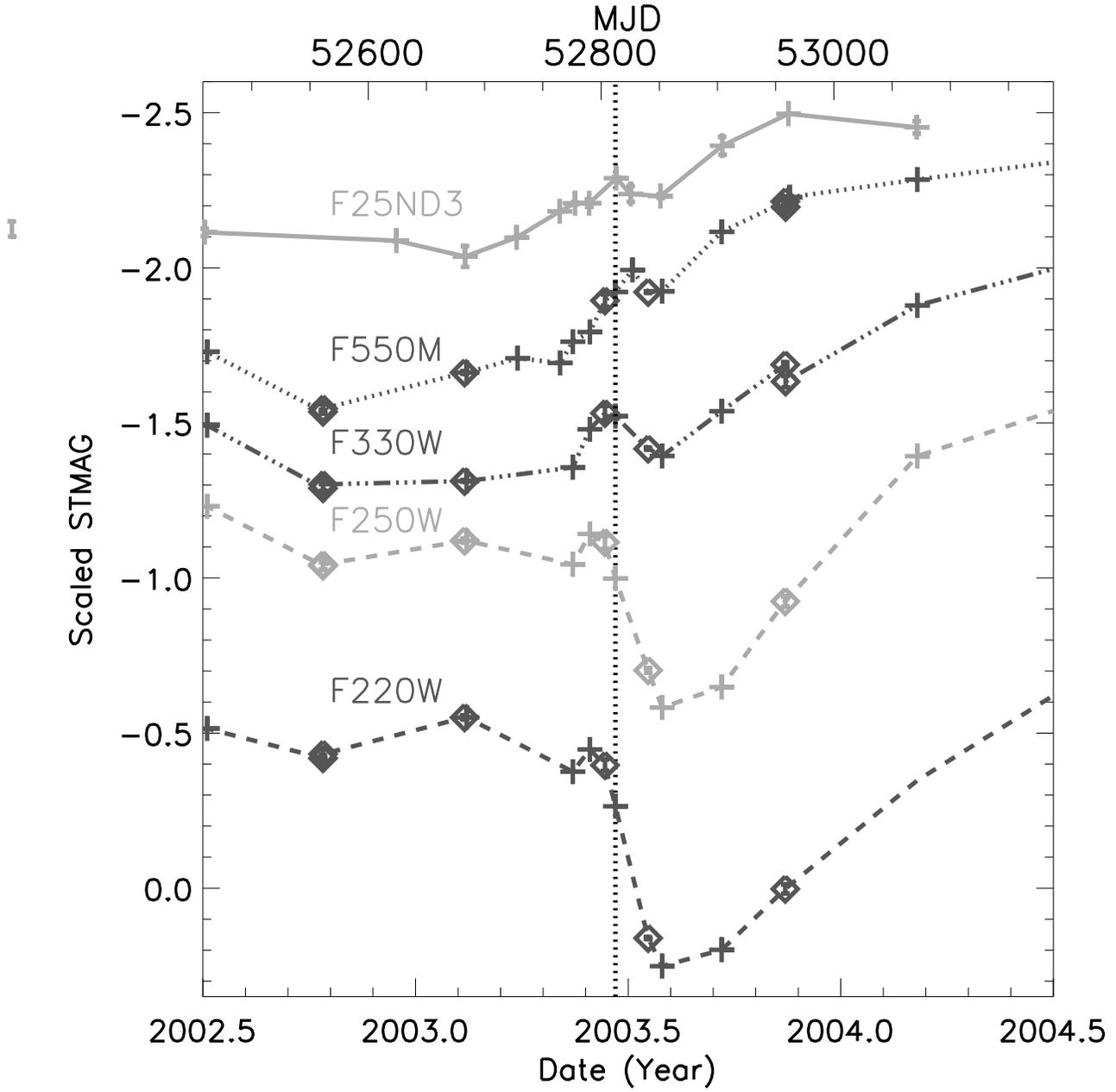}
\caption{Same as Fig \ref{plot1} with the time axis expanded around the 
spectroscopic event in 2003.}
\label{plot2}
\end{figure}

Figs.\ \ref{plot1} and \ref{plot2} reveal no major surprises 
about the 2003.5 spectroscopic event, but several comments are 
worthwhile.  First, the sharp drop in UV brightness (filters
F220W and F250W) is qualitatively understood and does not
involve circumstellar dust.  During both the 1998 and the 2003 
events, STIS data showed very heavy ultraviolet blanketing by 
ionized metal lines;  indeed the star became quite dark at some 
wavelengths between 2000 and 3000 {\AA} \citep{ironcurtain}.
We further note that just before the spectroscopic event, a 
slight increase occurred at wavelengths below 4000 {\AA}
(filters F220W, F250W, F330W), but not at visual and far-red
wavelengths (F550M and F25ND3).  Ground-based visual-wavelength
and near-IR photometry showed a qualitatively similar effect
\citep{vangenderen03,whitelock04,laplata}.  The brightening is particularly
prominent in J, H, and K which are dominated by free-free emission
\citep{whitelock04}. The ACS F550M and STIS F25ND3 data primarily
measure the continuum brightness, while the other HST filters are
heavily influenced by strong emission or absorption lines.    At
about the same time \ion{He}{1} emission in the central star also went
through a similar increase in brightness \citep{martinconf}.  The minor
pre-event brightening thus appears to represent an increase in some
emission features implying a temporary increase in ionizing flux.  The
primary star may provide additional UV photons or the hypothetical hot
companion star may excite the primary wind more than usual at that
time (just before periastron), but no quantitative model has been
attempted. 

Figs.\ \ref{plot1} and \ref{plot2} contain interesting hints  
about the timescale for the star's post-event recovery.  Four 
months after the 2003.5 event, for instance, the 2--10 keV 
X-ray flux had increased almost to a normal level \citep{xray}. 
The HST/ACS F220W and F250W brightnesses, however, were still 
quite low at that time, and they required about eight months 
to recover.  This timescale must be explained in any valid model 
for the spectroscopic events. 

\citet{halpha05} noted serious differences between STIS
spectra of the 1998.0 and 2003.5 events, and interpreted them
as evidence for a rapid secular physical change in $\eta$ Car.
\citet{damineli99} had earlier found that \ion{He}{1} emission 
became progressively weaker after each of the last few 
spectroscopic events.  These clues are obviously pertinent 
to our comments in Section \ref{stellarparams} above. 

Fluctuations {\it between\/} spectroscopic events have received
little attention in the past.   For instance, Fig.\ \ref{plot1} 
shows a brief 0.2-magnitude brightening at 2001.3; measured by the
STIS in both imaging and spectroscopic mode.  It was correlated 
with the behavior of a strange unidentified emission line 
near 6307 {\AA}, and with other subtle changes in the spectrum 
\citep{uemit}.  This is interesting because mid-cycle events
have not been predicted in any of the competing scenarios
for the 5.5-year cycle.  Perhaps the effects seen in 2001
indicate the level of basic, LBV-like fluctuations in
$\eta$ Car.  

\subsection{Eta Carinae in the Near Future \label{predictions}}  

The appearance of $\eta$ Car and the Homunculus nebula has
changed dramatically.  Twenty years ago the entire object  
could have been described as ``a bright, compact nebula having 
an indistinct eighth-magnitude central core;''   but a few years 
in the future, if recent trends continue, it will be seen 
instead as ``a fifth- or even fourth-magnitude star accompanied 
by some visible nebulosity.''  Meanwhile the color is gradually
becoming bluer.  This overall development has long been expected 
\citep{kd87}, but now appears to be occurring 20 years ahead of 
schedule.   If it signals an irregularity in the star's recovery 
from the Great Eruption, then this may be a highly unusual clue 
to the highly abnormal internal structure.  Unsteady diffusion of
either the thermal or the rotational parameters would be significant
for stellar astrophysics in general.  

There are several practical implications for future observations of 
this object.  Valid ground-based spectroscopy of the star (strictly 
speaking its wind) is becoming feasible for the first time, 
as its increased brightness overwhelms the emission-line contamination 
by inner ejecta.  Unfortunately this implies that the inner ejecta 
-- particularly the mysterious Weigelt knots -- are becoming difficult 
to observe.  In fact, since the HST/STIS is no longer available,
they are now practically impossible to observe.  When some new 
high-spatial-resolution spectrograph becomes available in the
future, the inner ejecta will probably be much fainter than the 
star.    

The expected future of the larger-scale Homunculus nebula is 
also interesting.  At present it is essentially a reflection
nebula.  However, based on the presence of high-excitation
emission lines such as [\ion{Ne}{3}] close to the star, the
system almost certainly contains a source of hydrogen-ionizing
photons with energies above 13.6, and helium-ionizing photons
above 25 eV.  (See, e.g., \citet{zanella84};  most recent 
authors assume that this source 
is a hot companion star.)  Eventually, when circumstellar 
extinction has decreased sufficiently due to expansion and other
effects, the UV source will begin to photoionize the  
Homunculus.  This is especially true if the primary 
stellar wind is weakening as we conjectured above.   
First the inner ``Little Homunculus'' will become a
bright compact \ion{H}{2} region, and then the bipolar
Homunculus lobes.  The time when that will occur is not
obvious, but it may be within the next few decades if
current trends continue.

\section{Acknowledgments}
This research was conducted as part of the HST Treasury Project on
$\eta$ Carinae via grant no. GO-9973 from the Space Telescope Science
Institute.   
We are grateful to T.R. Gull and Beth Perriello for
assisting with the HST observing plans.  
We also especially thank
Roberta Humphreys, J.T. Olds, and Matt Gray at the University of
Minnesota for discussions and helping with non-routine steps in the
data preparation and analysis.





\clearpage


\clearpage

\begin{deluxetable}{lllrrrrrrr}
\tablewidth{0pt}
\tabletypesize{\tiny}
\tablecaption{Results from STIS Acquisition Images \label{acqdata}}
\tablecolumns{8}
\tablehead{
\colhead{Dataset}&\colhead{MJD}&\colhead{Year}&
\colhead{Flux\tablenotemark{c}}&
\colhead{Magnitude\tablenotemark{a}}&
\colhead{Average\tablenotemark{b}}&
\colhead{$\sigma$\tablenotemark{b}}&
}
\startdata
o8ma93xjq&53070.199&2004.177&9.753E-12&-0.56&-0.54&0.02\\
o8ma93hjq&53071.230&2004.180&9.403E-12&-0.52&\nodata&\nodata\\

\enddata
\tablenotetext{a}{Relative STIS magnitude in the F25ND3 filter zeroed on 1999.140}
\tablenotetext{b}{The average and sigma of individual measurements within one day of each other.  These values are plotted in Fig. 4.}
\tablenotetext{c}{Brightness measured in F25ND3 filter given as STIS flux units (erg/cm$^2$/s/\mbox{\AA}).}
\end{deluxetable}

\begin{deluxetable}{lcccc}
\tablewidth{0pt}
\tabletypesize{\tiny}
\tablecaption{Calibration Values for ACS/HRC Data\label{acscaltab}}
\tablecolumns{5}
\tablehead{
&\colhead{F220W}&\colhead{F250W}&\colhead{F330W}&\colhead{F550M}
}
\startdata
ACS/HRC Aperture Correction\tablenotemark{a}&
0.593$\pm$0.014&0.594$\pm$0.013&0.625$\pm$0.001&0.619$\pm$0.022\\
ACS/HRC Effective Bandwidth\tablenotemark{b}&
187.29&239.41&173.75&165.20\\
STIS Throughput Correction\tablenotemark{c}&
1.6318&1.2903&0.6003&0.5601\\
\enddata
\tablenotetext{a}{Ratio of the ACS/HRC of flux observed in 0.3$\arcsec$ weighted aperture to flux with an infinite aperture.}
\tablenotetext{b}{{\AA}, includes the ACS/HRC CCD response.}
\tablenotetext{c}{The ratio of expected throughput for the aperture
  used to measure the ACS/HRC photometry to the actual
  integrated throughput of the STIS slit in the same filter.} 
\end{deluxetable}

\begin{deluxetable}{lllrrrr}
\tablewidth{0pt}
\tabletypesize{\tiny}
\tablecaption{Results from ACS/HRC Images\label{acsdata}}
\tablecolumns{7}
\tablehead{
&&&\colhead{Exp Time}&\colhead{Flux}\\
\colhead{Dataset}&\colhead{MJD}&\colhead{Year}&\colhead{(sec)}&\colhead{Density\tablenotemark{c}}&
\colhead{Magnitude\tablenotemark{a}}&\colhead{Average\tablenotemark{b}}
}
\startdata
\multicolumn{7}{c}{HRC/F220W Filter}\\
\tableline\\
j8ma7ac7q&53345.391&2004.931&5.0&0.392&7.417&7.410$\pm$0.017\\
j8ma7acbq&53345.395&2004.931&5.0&0.391&7.418&\nodata\\
j8ma7acfq&53345.398&2004.931&5.0&0.389&7.424&\nodata\\
j8ma7acrq&53345.434&2004.931&5.0&0.405&7.382&\nodata\\
j8ma8aorq&53565.266&2005.534&5.0&0.462&7.239&7.228$\pm$0.011\\
j8ma8ap4q&53565.301&2005.534&5.0&0.472&7.216&\nodata\\
j8ma9aetq&53680.328&2005.849&5.0&0.467&7.226&7.230$\pm$0.006\\
j8ma9af1q&53680.344&2005.849&5.0&0.462&7.239&\nodata\\
j8ma9afaq&53680.355&2005.849&5.0&0.465&7.231&\nodata\\
j8ma9afnq&53680.371&2005.849&5.0&0.468&7.225&\nodata\\
j9p602req&53951.121&2006.591&4.0&0.496&7.162&7.166$\pm$0.002\\
j9p602rhq&53951.125&2006.591&4.0&0.493&7.169&\nodata\\
j9p602rkq&53951.125&2006.591&4.0&0.494&7.165&\nodata\\
j9p602rnq&53951.129&2006.591&4.0&0.493&7.168&\nodata\\
\tableline\\
\multicolumn{7}{c}{HRC/F250W Filter}\\
\tableline\\
j8ma7ac8q&53345.391&2004.931&1.4&0.939&6.469&6.463$\pm$0.016\\
j8ma7accq&53345.395&2004.931&1.4&0.941&6.466&\nodata\\
j8ma7achq&53345.402&2004.931&1.4&0.928&6.481&\nodata\\
j8ma7actq&53345.438&2004.931&1.4&0.967&6.437&\nodata\\
j8ma8aouq&53565.273&2005.534&1.4&1.029&6.369&6.353$\pm$0.016\\
j8ma8ap7q&53565.309&2005.534&1.4&1.060&6.337&\nodata\\
j8ma9aewq&53680.336&2005.849&1.4&1.126&6.271&6.270$\pm$0.007\\
j8ma9af5q&53680.348&2005.849&1.4&1.119&6.277&\nodata\\
j8ma9afqq&53680.348&2005.849&1.4&1.137&6.261&\nodata\\
j9p601qnq&53951.051&2006.591&1.0&1.261&6.148&6.149$\pm$0.003\\
j9p601qqq&53951.055&2006.591&1.0&1.264&6.145&\nodata\\
j9p601qtq&53951.059&2006.591&1.0&1.258&6.151&\nodata\\
j9p601qwq&53951.063&2006.591&1.0&1.256&6.152&\nodata\\
\tableline\\
\multicolumn{7}{c}{HRC/F330W Filter}\\
\tableline\\
j8ma7ac9q&53345.391&2004.931&0.8&1.150&6.248&6.243$\pm$0.019\\
j8ma7acdq&53345.395&2004.931&0.8&1.140&6.258&\nodata\\
j8ma7aclq&53345.406&2004.931&0.8&1.142&6.256&\nodata\\
j8ma7acxq&53345.441&2004.931&0.8&1.190&6.211&\nodata\\
j8ma8aoyq&53565.277&2005.534&0.8&1.190&6.211&6.201$\pm$0.011\\
j8ma8apbq&53565.313&2005.534&0.8&1.213&6.190&\nodata\\
j8ma9aezq&53680.340&2005.849&0.8&1.440&6.004&6.002$\pm$0.004\\
j8ma9af8q&53680.352&2005.849&0.8&1.440&6.004&\nodata\\
j8ma9afhq&53680.367&2005.849&0.8&1.441&6.003&\nodata\\
j8ma9afuq&53680.383&2005.849&0.8&1.452&5.995&\nodata\\
j9p601qxq&53951.066&2006.591&0.2&1.671&5.843&5.850$\pm$0.010\\
j9p601qzq&53951.066&2006.591&0.2&1.688&5.831&\nodata\\
j9p601r0q&53951.066&2006.591&0.2&1.666&5.846&\nodata\\
j9p602roq&53951.129&2006.591&0.3&1.638&5.864&\nodata\\
j9p602rpq&53951.133&2006.591&0.3&1.656&5.852&\nodata\\
j9p602rqq&53951.133&2006.591&0.3&1.643&5.861&\nodata\\
j9p602rrq&53951.133&2006.591&0.3&1.653&5.854&\nodata\\
\tableline\\
\multicolumn{7}{c}{HRC/F550M Filter}\\
\tableline\\
j8ma7acaq&53345.395&2004.931&0.1&1.337&6.084&6.085$\pm$0.016\\
j8ma7aceq&53345.398&2004.931&0.1&1.324&6.095&\nodata\\
j8ma7acoq&53345.410&2004.931&0.1&1.318&6.100&\nodata\\
j8ma7ad0q&53345.445&2004.931&0.1&1.370&6.059&\nodata\\
j8ma8ap1q&53565.281&2005.534&0.1&1.342&6.081&6.077$\pm$0.004\\
j8ma8apeq&53565.281&2005.534&0.1&1.351&6.073&\nodata\\
j8ma9af0q&53680.340&2005.849&0.1&1.589&5.897&5.891$\pm$0.009\\
j8ma9af9q&53680.355&2005.849&0.1&1.582&5.902&\nodata\\
j8ma9afkq&53680.371&2005.849&0.1&1.606&5.886&\nodata\\
j8ma9afxq&53680.387&2005.849&0.1&1.614&5.880&\nodata\\
j9p602ryq&53951.145&2006.591&0.1&1.812&5.755&5.757$\pm$0.005\\
j9p602s0q&53951.145&2006.591&0.1&1.798&5.763&\nodata\\
j9p602s2q&53951.145&2006.591&0.1&1.819&5.750&\nodata\\
j9p602s4q&53951.148&2006.591&0.1&1.801&5.761&\nodata\\
\enddata
\tablenotetext{a}{Magnitude on the STMAG system.  These aperture corrections given in Table \ref{acscaltab} are applied to these magnitudes.}
\tablenotetext{b}{The average and sigma of individual measurements in
  a set of exposures taken within a day of each other.}
\tablenotetext{c}{STMAG flux units are 10$^{-11}$ erg/cm$^2$/s/\mbox{\AA}.}
\end{deluxetable}

\begin{deluxetable}{llrrrr}
\tablewidth{0pt}
\tabletypesize{\tiny}
\tablecaption{HST STIS Data\label{stisspectra}\tablenotemark{a}}
\tablecolumns{6}
\tablehead{
\colhead{Root}&&\colhead{Slit Angle}&&\colhead{Central
  $\lambda$}&\colhead{Exp Length}\\
\colhead{Name}&\colhead{MJD}&\colhead{(deg)\tablenotemark{a}}
&\colhead{Grating}&\colhead{({\AA})}&\colhead{(sec)}
}
\startdata
o4j8010y0&50891.6& -28&G230MB&1854&456.0\\
o4j8010o0&50891.6& -28&G230MB&1995&360.0\\
o4j8011b0&50891.7& -28&G230MB&2135&180.0\\
o4j8011c0&50891.7& -28&G230MB&2276&180.0\\
o4j8011a0&50891.7& -28&G230MB&2416&240.0\\
o4j801040&50891.4& -28&G230MB&2557&300.0\\
o4j8010e0&50891.5& -28&G230MB&2697&290.0\\
o4j801050&50891.4& -28&G230MB&2836&330.0\\
o4j8010f0&50891.5& -28&G230MB&2976&348.0\\
o4j8010g0&50891.5& -28&G230MB&3115&336.0\\
o4j8010h0&50891.5& -28&G430M&3165&144.0\\
o4j801170&50891.7& -28&G430M&3423& 60.0\\
o4j801180&50891.7& -28&G430M&3680& 72.0\\
o4j8010z0&50891.6& -28&G430M&3936& 72.0\\
o4j801060&50891.4& -28&G430M&4194& 36.0\\
o4j8010x0&50891.6& -28&G430M&4961& 36.0\\
o4j8010i0&50891.5& -28&G430M&5216& 36.0\\
o4j801190&50891.7& -28&G430M&5471& 36.0\\
o4j8010d0&50891.5& -28&G750M&5734& 15.0\\
o4j801120&50891.7& -28&G750M&6252&  9.4\\
o556020t0&51230.6& -28&G230MB&1854&380.0\\
o556020k0&51230.5& -28&G230MB&1995&278.0\\
o55602110&51230.7& -28&G230MB&2135&150.0\\
o55602120&51230.7& -28&G230MB&2276&150.0\\
o55602100&51230.7& -28&G230MB&2416&200.0\\
o556020b0&51230.5& -28&G230MB&2557&300.0\\
o556020e0&51230.5& -28&G230MB&2697&280.0\\
o55602040&51230.5& -28&G230MB&2836&307.0\\
o556020z0&51230.6& -28&G230MB&2976&360.0\\
o556020c0&51230.5& -28&G230MB&3115&280.0\\
o556020v0&51230.6& -28&G430M&3165&120.0\\
o556020f0&51230.5& -28&G430M&3423& 50.0\\
o556020d0&51230.5& -28&G430M&3680& 60.0\\
o556020s0&51230.6& -28&G430M&3936& 60.0\\
o556020y0&51230.6& -28&G430M&4961& 30.0\\
o556020g0&51230.5& -28&G430M&5216& 30.0\\
o556020w0&51230.6& -28&G430M&5471& 30.0\\
o55602090&51230.5& -28&G750M&5734& 15.0\\
o556020p0&51230.6& -28&G750M&6252&  8.0\\
o62r010p0&52016.9&  22&G230MB&1854&340.0\\
o62r010i0&52016.8&  22&G230MB&1995&300.0\\
o62r010m0&52016.8&  22&G230MB&2135&280.0\\
o62r010x0&52016.9&  22&G230MB&2276&300.0\\
o62r010y0&52016.9&  22&G230MB&2416&200.0\\
o62r010a0&52016.8&  22&G230MB&2557&420.0\\
o62r010c0&52016.8&  22&G230MB&2697&280.0\\
o62r01040&52016.8&  22&G230MB&2836&300.0\\
o62r010t0&52016.9&  22&G230MB&2976&360.0\\
o62r01080&52016.8&  22&G230MB&3115&350.0\\
o62r010r0&52016.9&  22&G430M&3165&100.0\\
o62r010d0&52016.8&  22&G430M&3423& 70.0\\
o62r010b0&52016.8&  22&G430M&3680& 60.0\\
o62r010o0&52016.8&  22&G430M&3936& 32.0\\
o62r01030&52016.7&  22&G430M&4194& 30.0\\
o62r010v0&52016.9&  22&G430M&4961& 36.0\\
o62r010e0&52016.8&  22&G430M&5216& 14.0\\
o62r010s0&52016.9&  22&G430M&5471& 30.0\\
o62r01090&52016.8&  22&G750M&5734&  8.0\\
o62r010l0&52016.8&  22&G750M&6252& 10.0\\
o6ex030e0&52183.2& 165&G430M&4961& 36.0\\
o6ex030c0&52183.1& 165&G430M&5216& 36.0\\
o6ex030b0&52183.1& 165&G750M&5734& 15.0\\
o6ex02080&52294.0& -82&G230MB&1854&800.0\\
o6ex020i0&52294.1& -82&G230MB&1995&600.0\\
o6ex020m0&52294.1& -82&G230MB&2135&600.0\\
o6ex020x0&52294.2& -82&G230MB&2276&600.0\\
o6ex020y0&52294.2& -82&G230MB&2416&320.0\\
o6ex020a0&52294.0& -82&G230MB&2557&1200.0\\
o6ex020c0&52294.1& -82&G230MB&2697&280.0\\
o6ex02040&52294.0& -82&G230MB&2836&300.0\\
o6ex020t0&52294.1& -82&G230MB&2976&340.0\\
o6ex020p0&52294.1& -82&G230MB&3115&300.0\\
o6ex020r0&52294.1& -82&G430M&3165& 90.0\\
o6ex020d0&52294.1& -82&G430M&3423& 90.0\\
o6ex020b0&52294.1& -82&G430M&3680& 52.0\\
o6ex020o0&52294.1& -82&G430M&3936& 26.0\\
o6ex02030&52294.0& -82&G430M&4194& 18.0\\
o6ex020v0&52294.2& -82&G430M&4961& 36.0\\
o6ex020e0&52294.1& -82&G430M&5216& 16.0\\
o6ex020s0&52294.1& -82&G430M&5471& 34.0\\
o6ex02090&52294.0& -82&G750M&5734&  6.0\\
o6ex020l0&52294.1& -82&G750M&6252&  8.0\\
o6mo020a0&52459.5&  69&G230MB&1854&400.0\\
o6mo020x0&52459.6&  69&G230MB&1995&300.0\\
o6mo02120&52459.6&  69&G230MB&2135&300.0\\
o6mo021n0&52459.7&  69&G230MB&2276&300.0\\
o6mo021e0&52459.7&  69&G230MB&2416&320.0\\
o6mo020h0&52459.6&  69&G230MB&2557&400.0\\
o6mo020m0&52459.6&  69&G230MB&2697&340.0\\
o6mo02050&52459.5&  69&G230MB&2836&300.0\\
o6mo021i0&52459.7&  69&G230MB&2976&320.0\\
o6mo02190&52459.7&  69&G230MB&3115&300.0\\
o6mo021r0&52459.7&  69&G430M&3165& 90.0\\
o6mo020p0&52459.6&  69&G430M&3423& 90.0\\
o6mo020l0&52459.6&  69&G430M&3680& 52.0\\
o6mo021a0&52459.7&  69&G430M&3936& 26.0\\
o6mo02060&52459.5&  69&G430M&4194& 18.0\\
o6mo021m0&52459.7&  69&G430M&4961& 36.0\\
o6mo020q0&52459.6&  69&G430M&5216& 16.0\\
o6mo021h0&52459.7&  69&G430M&5471& 34.0\\
o6mo020i0&52459.6&  69&G750M&5734&  9.0\\
o6mo02150&52459.7&  69&G750M&6252&  8.0\\
o8gm12060&52682.9& -57&G230MB&1854&600.0\\
o8gm120h0&52683.0& -57&G230MB&1995&600.0\\
o8gm120l0&52683.0& -57&G230MB&2135&600.0\\
o8gm120w0&52683.0& -57&G230MB&2276&600.0\\
o8gm120r0&52683.0& -57&G230MB&2416&320.0\\
o8gm12090&52682.9& -57&G230MB&2557&800.0\\
o8gm120c0&52682.9& -57&G230MB&2697&340.0\\
o8gm12030&52682.9& -57&G230MB&2836&300.0\\
o8gm120t0&52683.0& -57&G230MB&2976&340.0\\
o8gm120o0&52683.0& -57&G230MB&3115&300.0\\
o8gm120n0&52683.0& -57&G430M&3165& 90.0\\
o8gm120d0&52682.9& -57&G430M&3423& 90.0\\
o8gm120b0&52682.9& -57&G430M&3680& 52.0\\
o8gm120p0&52683.0& -57&G430M&3936& 26.0\\
o8gm12040&52682.9& -57&G430M&4194& 18.0\\
o8gm120v0&52683.0& -57&G430M&4961& 36.0\\
o8gm120e0&52682.9& -57&G430M&5216& 16.0\\
o8gm120s0&52683.0& -57&G430M&5471& 34.0\\
o8gm120a0&52682.9& -57&G750M&5734&  6.0\\
o8gm12050&52682.9& -57&G750M&6252&  8.0\\
o8gm210g0&52727.3& -28&G430M&4961& 16.0\\
o8gm210e0&52727.3& -28&G430M&5216& 16.0\\
o8gm210b0&52727.3& -28&G430M&5471& 34.0\\
o8gm210d0&52727.3& -28&G750M&5734&  9.0\\
o8gm410g0&52764.4&  27&G430M&4961& 16.0\\
o8gm410e0&52764.4&  27&G430M&5216& 16.0\\
o8gm410b0&52764.3&  27&G430M&5471& 34.0\\
o8gm410d0&52764.3&  27&G750M&5734&  9.0\\
o8gm320a0&52778.5&  38&G230MB&1854&400.0\\
o8gm320x0&52778.9&  38&G230MB&1995&300.0\\
o8gm33020&52776.4&  38&G230MB&2135&300.0\\
o8gm330n0&52776.6&  38&G230MB&2276&300.0\\
o8gm330e0&52776.5&  38&G230MB&2416&320.0\\
o8gm320h0&52778.6&  38&G230MB&2557&400.0\\
o8gm320m0&52778.7&  38&G230MB&2697&340.0\\
o8gm32050&52778.5&  38&G230MB&2836&300.0\\
o8gm330i0&52776.5&  38&G230MB&2976&320.0\\
o8gm33090&52776.5&  38&G230MB&3115&300.0\\
o8gm33060&52776.4&  38&G430M&3165& 90.0\\
o8gm320p0&52778.8&  38&G430M&3423& 90.0\\
o8gm320l0&52778.7&  38&G430M&3680& 52.0\\
o8gm330a0&52776.5&  38&G430M&3936& 26.0\\
o8gm32060&52778.5&  38&G430M&4194& 18.0\\
o8gm330m0&52776.6&  38&G430M&4961& 36.0\\
o8gm320q0&52778.8&  38&G430M&5216& 16.0\\
o8gm330h0&52776.5&  38&G430M&5471& 34.0\\
o8gm320i0&52778.6&  38&G750M&5734&  9.0\\
o8gm330r0&52776.6&  38&G750M&6252&  8.0\\
o8gm520a0&52791.7&  62&G230MB&1854&400.0\\
o8gm520x0&52791.8&  62&G230MB&1995&300.0\\
o8gm52100&52791.9&  62&G230MB&2135&400.0\\
o8gm521o0&52792.0&  62&G230MB&2276&300.0\\
o8gm521f0&52791.9&  62&G230MB&2416&320.0\\
o8gm520h0&52791.8&  62&G230MB&2557&400.0\\
o8gm520m0&52791.8&  62&G230MB&2697&340.0\\
o8gm52050&52791.7&  62&G230MB&2836&300.0\\
o8gm521j0&52791.9&  62&G230MB&2976&340.0\\
o8gm52180&52791.9&  62&G230MB&3115&300.0\\
o8gm52170&52791.9&  62&G430M&3165& 90.0\\
o8gm520p0&52791.8&  62&G430M&3423& 90.0\\
o8gm520l0&52791.8&  62&G430M&3680& 52.0\\
o8gm521b0&52791.9&  62&G430M&3936& 26.0\\
o8gm52060&52791.7&  62&G430M&4194& 18.0\\
o8gm521k0&52791.9&  62&G430M&4961& 36.0\\
o8gm520q0&52791.8&  62&G430M&5216& 16.0\\
o8gm521g0&52791.9&  62&G430M&5471& 34.0\\
o8gm520i0&52791.8&  62&G750M&5734&  9.0\\
o8gm521s0&52792.0&  62&G750M&6252&  8.0\\
o8gm620a0&52813.8&  70&G230MB&1854&400.0\\
o8gm620x0&52814.2&  70&G230MB&1995&300.0\\
o8gm62100&52814.2&  70&G230MB&2135&300.0\\
o8gm630d0&52812.2&  70&G230MB&2276&300.0\\
o8gm63040&52812.1&  70&G230MB&2416&350.0\\
o8gm620h0&52814.0&  70&G230MB&2557&400.0\\
o8gm620m0&52814.1&  70&G230MB&2697&340.0\\
o8gm62050&52813.8&  70&G230MB&2836&300.0\\
o8gm63080&52812.2&  70&G230MB&2976&320.0\\
o8gm62170&52814.3&  70&G230MB&3115&260.0\\
o8gm62140&52814.3&  70&G430M&3165& 90.0\\
o8gm620p0&52814.1&  70&G430M&3423& 90.0\\
o8gm620l0&52814.1&  70&G430M&3680& 52.0\\
o8gm62180&52814.3&  70&G430M&3936& 26.0\\
o8gm62060&52813.8&  70&G430M&4194& 18.0\\
o8gm630c0&52812.2&  70&G430M&4961& 36.0\\
o8gm620q0&52814.1&  70&G430M&5216& 16.0\\
o8gm63070&52812.1&  70&G430M&5471& 34.0\\
o8gm620i0&52814.0&  70&G750M&5734&  9.0\\
o8gm630h0&52812.2&  70&G750M&6252& 15.0\\
o8ma720q0&52825.5&  69&G430M&4961& 16.0\\
o8ma720p0&52825.5&  69&G430M&5216& 16.0\\
o8ma720h0&52825.4&  69&G750M&5734&  9.0\\
o8ma820b0&52852.0& 105&G230MB&1854&400.0\\
o8ma820y0&52852.2& 105&G230MB&1995&300.0\\
o8ma82110&52852.2& 105&G230MB&2135&300.0\\
o8ma821m0&52852.4& 105&G230MB&2276&300.0\\
o8ma821b0&52852.3& 105&G230MB&2416&320.0\\
o8ma820i0&52852.0& 105&G230MB&2557&400.0\\
o8ma820n0&52852.1& 105&G230MB&2697&340.0\\
o8ma82060&52851.9& 105&G230MB&2836&300.0\\
o8ma821i0&52852.4& 105&G230MB&2976&300.0\\
o8ma82160&52852.3& 105&G230MB&3115&300.0\\
o8ma821a0&52852.3& 105&G430M&3165& 90.0\\
o8ma820q0&52852.1& 105&G430M&3423& 90.0\\
o8ma820m0&52852.1& 105&G430M&3680& 52.0\\
o8ma821e0&52852.3& 105&G430M&3936& 26.0\\
o8ma82070&52851.9& 105&G430M&4194& 18.0\\
o8ma821o0&52852.4& 105&G430M&4961& 32.0\\
o8ma820r0&52852.1& 105&G430M&5216& 16.0\\
o8ma821j0&52852.4& 105&G430M&5471& 34.0\\
o8ma820j0&52852.1& 105&G750M&5734&  6.0\\
o8ma820a0&52852.0& 105&G750M&6252&  8.0\\
o8ma92070&52904.3& 153&G230MB&1854&600.0\\
o8ma920i0&52904.4& 153&G230MB&1995&600.0\\
o8ma920m0&52904.4& 153&G230MB&2135&600.0\\
o8ma920x0&52904.5& 153&G230MB&2276&300.0\\
o8ma920s0&52904.4& 153&G230MB&2416&600.0\\
o8ma920a0&52904.3& 153&G230MB&2557&800.0\\
o8ma920d0&52904.4& 153&G230MB&2697&340.0\\
o8ma92040&52904.3& 153&G230MB&2836&300.0\\
o8ma920u0&52904.5& 153&G230MB&2976&340.0\\
o8ma920p0&52904.4& 153&G230MB&3115&300.0\\
o8ma920o0&52904.4& 153&G430M&3305& 90.0\\
o8ma920e0&52904.4& 153&G430M&3423& 90.0\\
o8ma920c0&52904.4& 153&G430M&3680& 52.0\\
o8ma920q0&52904.4& 153&G430M&3936& 26.0\\
o8ma92050&52904.3& 153&G430M&4194& 18.0\\
o8ma920w0&52904.5& 153&G430M&4961& 36.0\\
o8ma920f0&52904.4& 153&G430M&5216& 16.0\\
o8ma920t0&52904.5& 153&G430M&5471& 34.0\\
o8ma920b0&52904.4& 153&G750M&5734&  6.0\\
o8ma920z0&52904.5& 153&G750M&6252&  8.0\\
o8ma830g0&52960.7&-142&G430M&4961& 18.0\\
o8ma830e0&52960.6&-142&G430M&5216& 16.0\\
o8ma830b0&52960.6&-142&G430M&5471& 32.0\\
o8ma830d0&52960.6&-142&G750M&5734&  8.0\\
o8ma940g0&53071.3& -28&G230MB&1854&430.0\\
o8ma940i0&53071.3& -28&G230MB&2135&320.0\\
o8ma940m0&53071.3& -28&G230MB&2416&450.0\\
o8ma94070&53071.3& -28&G230MB&2557&410.0\\
o8ma940e0&53071.3& -28&G230MB&2697&323.0\\
o8ma94020&53071.2& -28&G230MB&2836&320.0\\
o8ma940s0&53071.3& -28&G230MB&2976&323.0\\
o8ma940j0&53071.3& -28&G230MB&3115&255.0\\
o8ma940h0&53071.3& -28&G430M&3165& 90.0\\
o8ma940a0&53071.3& -28&G430M&3423& 90.0\\
o8ma94090&53071.3& -28&G430M&3680& 52.0\\
o8ma940k0&53071.3& -28&G430M&3936& 26.0\\
o8ma94030&53071.2& -28&G430M&4194& 18.0\\
o8ma940p0&53071.3& -28&G430M&4961& 36.0\\
o8ma940b0&53071.3& -28&G430M&5216& 16.0\\
o8ma940n0&53071.3& -28&G430M&5471& 34.0\\
o8ma94080&53071.3& -28&G750M&5734&  6.0\\
o8ma940r0&53071.3& -28&G750M&6252& 10.0\\
\enddata
\tablenotetext{a}{The slit angle is measured from north through east.
  All slits are peaked up on the central star.  The 52\arcsec x
  0.1\arcsec slit was used for all these observations.}
\end{deluxetable}

\begin{deluxetable}{rrrr}
\tablewidth{0pt}
\tabletypesize{\tiny}
\tablecaption{Synthetic HST STIS Photometry\label{fakeacs}}
\tablecolumns{4}
\tablehead{
&&\colhead{Flux}\\
\colhead{MJD}&\colhead{Year}&\colhead{Density\tablenotemark{a}}&\colhead{STMAG\tablenotemark{a}}
}
\startdata
\multicolumn{4}{c}{HRC/F220W Filter}\\
\tableline\\
50891.7&1998.21&0.076&9.194\\
51230.6&1999.14&0.155&8.425\\
52016.9&2001.29&0.189&8.211\\
52294.2&2002.05&0.230&7.996\\
53459.7&2002.51&0.255&7.885\\
52683.0&2003.12&0.263&7.850\\
52777.6&2003.37&0.224&8.025\\
52791.9&2003.41&0.239&7.953\\
52813.1&2003.47&0.202&8.136\\
52852.2&2003.58&0.126&8.652\\
52904.4&2003.72&0.132&8.599\\
53071.3&2004.18&0.218&8.053\\
\tableline\\
\multicolumn{4}{c}{HRC/F250W Filter}\\
\tableline\\
50891.6&1998.21&0.167&8.346\\
51230.6&1999.14&0.403&7.387\\
52016.9&2001.29&0.448&7.273\\
52294.2&2002.05&0.550&7.048\\
53459.7&2002.51&0.592&6.969\\
52683.0&2003.12&0.535&7.080\\
52777.6&2003.37&0.498&7.156\\
52791.9&2003.41&0.546&7.058\\
52813.1&2003.47&0.478&7.202\\
52852.2&2003.58&0.326&7.618\\
52904.4&2003.72&0.346&7.552\\
53071.3&2004.18&0.687&6.807\\
\tableline\\
\multicolumn{4}{c}{HRC/F330W Filter}\\
\tableline\\
50891.6&1998.21&0.237&7.964\\
51230.6&1999.14&0.516&7.119\\
52016.9&2001.29&0.524&7.101\\
52294.1&2002.05&0.625&6.910\\
53459.7&2002.51&0.626&6.908\\
52683.0&2003.12&0.531&7.088\\
52777.6&2003.37&0.553&7.043\\
52791.9&2003.41&0.619&6.920\\
52813.1&2003.47&0.644&6.878\\
52852.2&2003.58&0.572&7.007\\
52904.4&2003.72&0.653&6.862\\
53071.3&2004.18&0.894&6.521\\
\tableline\\
\multicolumn{4}{c}{HRC/F550M Filter}\\
\tableline\\
50891.6&1998.21&0.352&7.533\\
51230.6&1999.14&0.605&6.945\\
52016.9&2001.29&0.691&6.802\\
52183.2&2001.75&0.823&6.611\\
52294.1&2002.05&0.703&6.782\\
53459.7&2002.51&0.711&6.771\\
52683.0&2003.12&0.668&6.838\\
52727.3&2003.24&0.697&6.792\\
52764.4&2003.34&0.688&6.807\\
52777.6&2003.37&0.732&6.738\\
52791.9&2003.41&0.754&6.707\\
52813.1&2003.47&0.849&6.578\\
52825.5&2003.51&0.906&6.507\\
52852.2&2003.58&0.851&6.575\\
52904.5&2003.72&1.015&6.384\\
52960.6&2003.88&1.124&6.273\\
53071.3&2004.18&1.185&6.215\\
\enddata
\tablenotetext{a}{Flux density in units of 10$^{-11}$ erg/cm$^2$/s/\mbox{\AA}.  The flux density and STMAG are corrected using the factors given in Table \ref{acscaltab}.}
\end{deluxetable}

\end{document}